\documentclass[manuscript]{aastex}
\usepackage{natbib}
\usepackage{graphicx}  
\bibpunct{(}{)}{;}{a}{}{,} % to follow the A&A style

\begin{document}
\renewcommand{\thefootnote}{\fnsymbol{footnote}}

  \title{Emission lines between 1 and 2 keV in Cometary X-ray Spectra}

\author{Ian Ewing, Damian J. Christian}
\affil{Department of Physics and Astronomy, California State University, 
18111 Nordhoff Street, Northridge, CA
91330; ian.ewing.794@my.csun.edu; daman.christian@csun.edu} 
\author{Dennis~Bodewits} 
\affil{Department of Astronomy, University of Maryland, College Park, MD 20742}
                  \author{Konrad~Dennerl}
         \affil{Max-Planck-Institut f\"{u}r extraterrestrische Physik, 
         Postfach 1312, 85741 Garching
         %Giessenbachstrasse, 85748 Garching, 
         Germany}
%            \email{kod@mpe.mpg.de}
\author{Carey M. Lisse}
         \affil{Planetary Exploration Group, Space Department, Johns Hopkins University Applied Physics Laboratory, 11100 Johns Hopkins Rd, Laurel, MD 20723, USA}
          \author{Scott~J.~Wolk}
           \affil{Harvard-Smithsonian Center for Astrophysics, 60 Garden Street, Cambridge, MA 02138, USA}
 %             \email{carey.lisse@jhuapl.edu}
%            \email{swolk@head.cfa.harvard.edu}
 % }
%   \date{Received \today}

\begin{abstract}
We present the detection of new cometary X-ray emission lines in the 1.0 to 2.0~keV range using a sample of comets observed with the \emph{Chandra} X-ray observatory
and \textsc{acis} spectrometer.
We have selected 5 comets from the \emph{Chandra} sample with good signal-to-noise spectra. 
The surveyed comets are: C/1999~S4~(\textsc{linear}), C/1999 T1 (McNaught--Hartley), 
153P/2002 (Ikeya--Zhang), 2P/2003 (Encke), and C/2008 8P (Tuttle).
We modeled the 
spectra with an extended version of our solar wind charge exchange (SWCX)  emission
model \citep{Bod07}.
Above 1 keV, we find Ikeya--Zhang to have  strong emission lines at 1340 and 1850 eV that we identify as being created by solar wind charge exchange lines of Mg XI and Si XIII, respectively, and weaker emission lines at 1470, 1600, and 1950 eV formed by SWCX of Mg XII, Mg XI, and Si~XIV, respectively. The Mg XI and XII and Si XIII and XIV lines are detected at a significant level for the other 
%and Si XIII, respectively.  The Mg XI 1340 and other Mg lines are significant for the other 
comets in our sample (LS4, MH, Encke, 8P),
and these lines promise additional diagnostics to be included in SWCX models.
%and promise additional diagnostic lines to be added to SWCX models.
The silicon lines in the 1700 to 2000 eV range are detected for all comets, but with the rising background and decreasing cometary emission, we caution these detections need further confirmation with higher resolution instruments. 

\end{abstract}

%   \maketitle
%\clearpage
%
%_____________________________________________________________________
\section{Introduction}

When highly charged ions from the solar wind collide on a neutral
gas, the ions get partially neutralized by capturing electrons
into an excited state. The subsequent line emission, called solar wind charge
exchange emission (SWCX\footnote[1]{Although there are many non-solar system examples of charge exchange, we use the SWCX abbreviation for the present work on comets.}) in X--ray and the Far--UV has been
observed in comets, planets, and the ISM. Please see \citet{Lis96, Kra97, Sno04, BEG07, Den10} for reviews.
 
The SWCX process has gained considerable interest recently, as it may 
contribute a significant amount %to the soft X-ray background [8,9], 
to the soft X-ray background \citep{Sno04}, %(Snowden et al. 2004), 
the interaction of the solar wind and the Earth's magnetosphere,
%[10,11], the solar heliosphere, and may be detectable from the 
\citep{Fuj07, W04}
%(Fujimoto et al. 2007; Wargelin et al. 2004) 
and may be detectable from the 
exospheres of other stars \citep{WD01}. % (Wargelin and Drake 2001). 
More recently SWCX has also been invoked to explain excess emission in the 2 $-$ 3 keV region from hot diffuse plasma found in galaxies \citep{Liu11}, %(Liu et al. 2011) 
star forming regions \citep{Tow11}, and from the Cygnus loop supernova remnant \citep{Kat11}, for example. %(Townsley et al. 2010).

Cometary SWCX models have evolved over the last several years, from the initial models of \citet{Hab97} and \citet{Cra97} for example,  to models considering the importance of the solar wind properties in emission 
\citep{SC00, KD00, KD01, Bei03, Kha03, Bod06}.
The recent survey of a sample of comets observed with \emph{Chandra} by \citet{Bod07}
has shown that the shape of cometary SWCX spectra is predominantly determined by the state of the incoming solar wind. SWCX spectra between 0.3 $-$ 1.0 keV are relatively well understood, aided by a large body of experimental work with hydrogen-like and fully stripped carbon and oxygen ions. This is not the case below
0.3 keV, where there is a forest of emission lines of elements such as Ne, Mg, Si etc. 
%\citep{SC00} that cannot be resolved,
\citep{SC00} that cannot be distinguished,
because of poor instrumental resolution. Observations of comet Schwassmann--Wachmann~3B (73P), a comet that passed within 0.07 AU
of Earth, allowing a direct comparison with in situ measurements of the composition of the solar wind, showed that 75 percent of the emission below 300~eV could not be accounted for by carbon ions alone \citep{W09}.

Similarly, the X-ray spectrum above 1 keV has not been well studied although tentative detections of neon ions, in the 1.1 -- 1.2 keV range, were made in the spectrum of the exceptionally X-ray bright comet Ikeya--Zhang presented in \citet{Bod07}.
%Carter et al. (2009)
\citet{Car10} found clear evidence of various Si and Mg, lines in XMM spectra, attributed to interaction of an interplanetary coronal mass ejection with the Earth's exosphere, and \citet{Fuj07} reported Mg emisson from the north ecliptic pole in {\it Suzaku} observations. Quantifying emission lines in this spectral range from comets is the goal of our current work.

%paper by section
Here, we present new line emission features discovered in \emph{Chandra} cometary X-ray spectra in the 1000 to 2000 eV range.
%charge exchange reactions!(and higher energy charge exchange components exist)!. 
We present details of the \emph{Chandra} X-ray observations, background subtraction methods, and the SWCX models used in our analysis in Section \ref{sec:obs}. 
Results for the X-ray spectra are presented in \ref{sec:results}.
In Section \ref{sec:discussion}, we discuss our results, line fluxes, line ratios and identifications, and compare these to previous X-ray observations. %  in terms of comet and solar wind characteristics. 
Lastly, in Section~5 we summarize our findings.

\section{Observations and Analysis}\label{sec:obs}

\subsection{\emph{Chandra} Observations}
Since its launch in 1999, 11 comets have been observed with the
\emph{Chandra} X-ray Observatory and Advanced \textsc{ccd} Imaging
Spectrometer (\textsc{acis}).  Here, we only include observations
made with the \textsc{acis}-S3 chip, which has the most sensitive low
energy response.  We have selected 5 comets from the \emph{Chandra} sample with good signal-to-noise spectra and the details of their observations are summarized in Table 1.

ACIS-S is a moderate resolution X-ray imaging CCD with a plate scale of 0.5$\arcsec$/pixel, an instantaneous field of view 8.3$\arcmin \times 8.3\arcmin$, and moderate resolution spectra ($\Delta E \sim$ 110 eV FWHM, $\sigma_{Gaussian} \sim$ 50 eV) in the 300 - 2000 eV energy range. The observations were conducted with the comet's nucleus near the aim-point in the S3 chip and no active guiding on the comet is used.   
In this "drift-scan" method, the comet is centered in the S3 chip and allowed to drift across the chip and the \emph{Chandra} pointing is updated to re-center the comet  before it moves off the chip edge. 

 \begin{figure}
   \centering
          \includegraphics[width=10.5cm, angle=90]{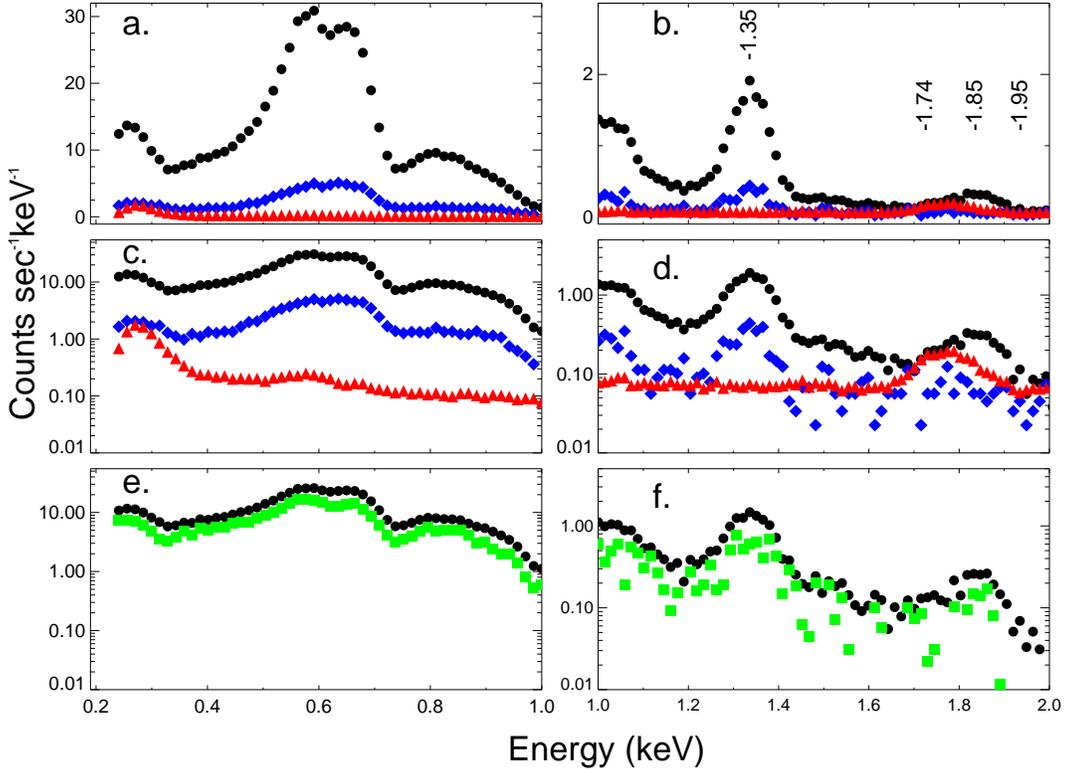} % Ian best line fit
      \caption{ACIS S3 Spectra of Ikeya--Zhang along with different backgrounds used.
Top panels show the ACIS S3 spectra on a linear scale with no background subtracted (filled circles) and over-plotted with the nominal S3 background in blue (diamonds) and the ACIS blank sky background in red (triangles) for: {\it a.}  the 0.2 -- 1.0 keV range and {\it b.} the 1.0 -- 2.0 keV range (some key energies are noted in panel b). Middle panels,  {\it c.} and {\it d.} are the same as {\it a.} and {\it b.}, but with a log scale for the ordinate. 
Shown in the bottom panels are IZ spectra with the nominal S3 background spectrum subtracted (filled circles) and over-plotted with the spectrum with 3 times that background subtracted in green (filled squares) for: {\it e.}  the 0.2 -- 1.0 keV range and {\it f.} the 1.0 -- 2.0 keV range.
       }
      \label{fig:iz_bg_4spec}
 \end{figure}

\begin{figure}[t]
   \centering
    \includegraphics[width=11.5cm]{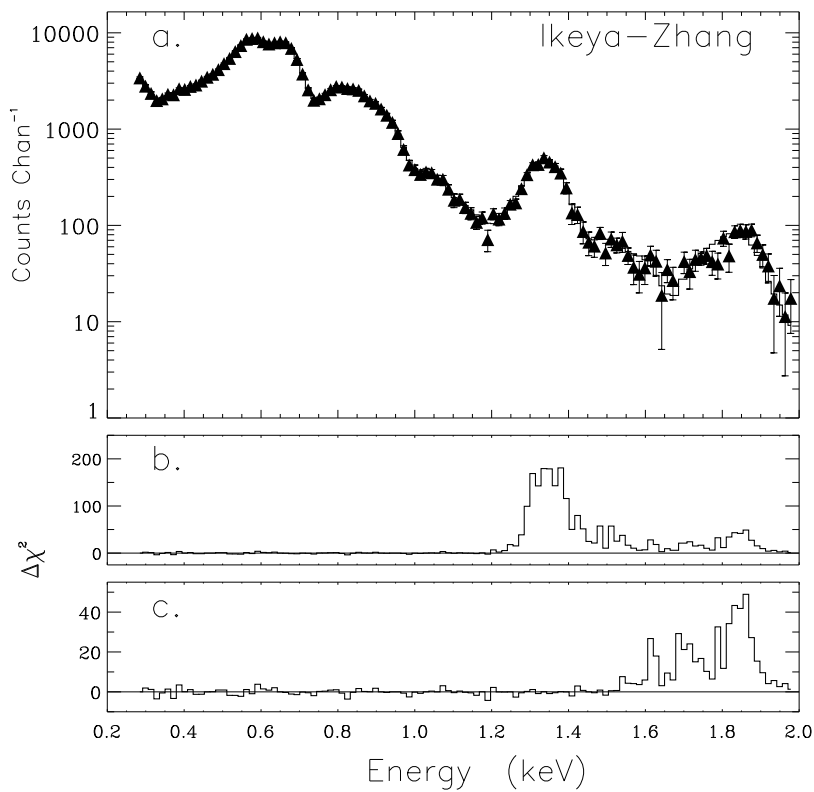}
      \caption{ACIS S3 Spectra of Ikeya--Zhang.
{\it a.} Top panel shows the data (with $\pm1\sigma$ error bars) and SWCX model fit as a solid histogram. 
The over-plotted model is our SWCX model %(Model 1)
including the addition of 
lines  at 1340, 1470, 1600, 1740, 1850, and 1950 eV (see text).
{\it b.} The {\it middle} panel shows the $\chi^2$ residuals of the data - model ($\Delta\chi^2$) with no additional lines above 1.3 keV, 
 and {\it c.} the {\it lower} panel, shows the $\chi^2$ residuals of the data - model now with the inclusion of lines at 1340 and 1470 eV lines, but no lines at higher energies showing the excess emission between 1600 and 2000 eV.
 }
         \label{fig:iz}
 \end{figure}

  \begin{figure}
   \centering
               \includegraphics[width=11.5cm]{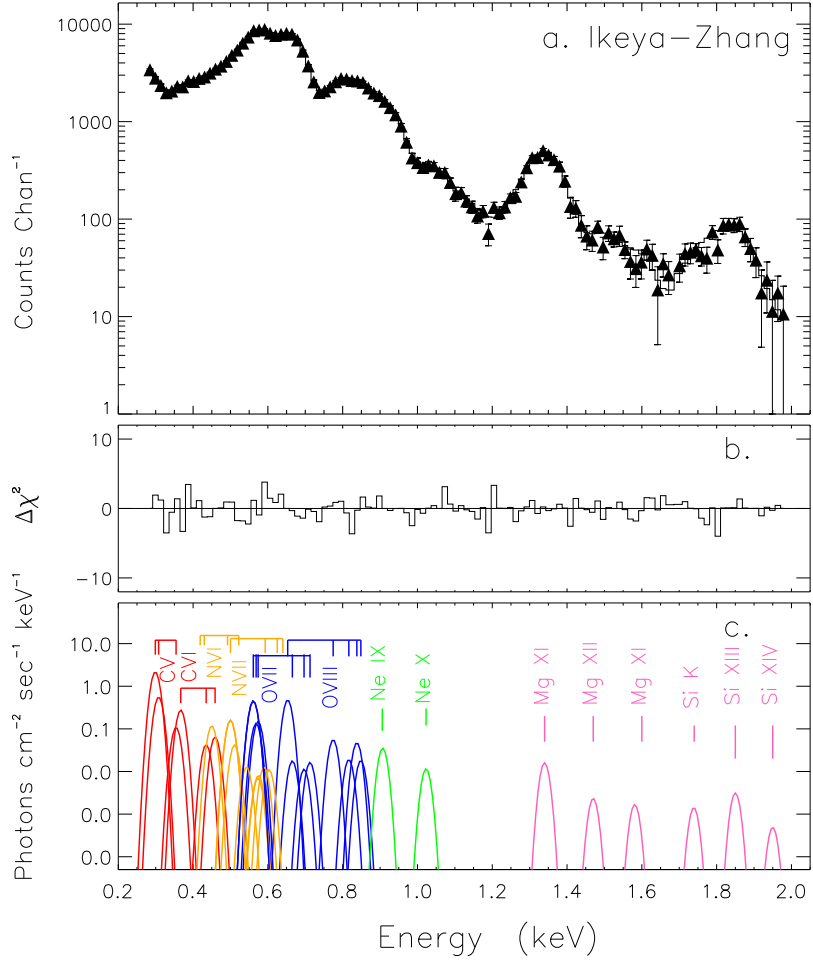} % CXE26 best 
      \caption{ACIS S3 Spectra of Ikeya--Zhang.
{\it a.} Shown in the top panel are the data (with $\pm1\sigma$ error bars) and SWCX model as a solid histogram.
{\it b.} The middle panel shows the $\chi^2$ residuals of the data - model ($\Delta\chi^2$).
{\it c.} The lower panel
shows the incident SWCX spectrum and the individual lines ($\sigma\approx$50eV)
for CV and CVI (red),
NV and NVI (orange), OVII and OVIII (blue) and Ne IX \& X (green) and new lines above 1.2 keV as indicated in pink. 
  }
         \label{fig:izmodel}
 \end{figure}

\begin{centering}
\begin{table}
%\begin{sidewaystable*}
%\begin{minipage}[t][180mm]{\textwidth}
\caption{Sample of \emph{Chandra} comet observations with observing parameters and solar wind data.} 
\begin{tabular}{l c c c c c c c c} \hline\hline
%\tablewidth{0pt}
\footnotesize
%\centering
Parameter\tablenotemark{a}  &   LS4\tablenotemark{b} &   MH &   IZ   &    Encke  & 8P   \\
\hline
Obs. Date       &   07/14     &   01/8-15      &   04/15-16     &   11/24   & 1/01-04  \\
       &    2000        &   2001      &  
 2002      &    2003   & 2008    \\
Prop. No. & 01100323 & 02100340 & 03108076 & 05100560 & 09100452 \\  
%Prop.No. & 1100323-LS$ & 2100340-MH & 03100792-WM1 & 3108076-IZ & 5100560 -Enck & 5108176-Q4 & 6100312-Tempel1 & 7108248-SW3 & 08108279-17P & 09100452-8P\\  
T$_{exp}$   (ks)  &   9.4    &   16.9    &   24      &   44   & 47      \\
r$_h$      (AU) &   0.8    &   1.26     &   0.81    &   0.89   & 1.10   \\
$\Delta$  (AU)  &   0.53   &   1.37   & 0.45      &   0.28   & 0.25     \\
%Lon, Lat   (Deg)   &   312, 24     &   185.3, 15.4     & 70,-34   &   206, 26.3       &   14.4,-9.6   & 211,-3 &   245, 0.79   &   247.2, 0.54 \\
%Phase   &  (Deg)   &           &   62.29       &  87      &   101.5       &   103 &   40.6    &       \\
%Radius FoV (km)    &           &   353 pix     &        &           &   550"    &       &       \\
Q$_{gas}$  &   3    &   6-20   & 20   &   0.7 & 2.2   \\
{\small (10$^{28}$~mols$^{-1}$)} & \\
v$_p$   &   592     &   353   &   372   &   583 & 360  \\
{\small (km s$^{-1})$} \\
F$_p$    &   2.9     &   1.6      &   3.8   &  3.2  & 0.2-0.5  \\
 (10$^8$ cm$^{-2}$ s$^{-1})$   \\
%$\Delta t$     (days)  &   1.09        &   6.63        &       &   -0.73       &   -1.09   &       &   0.2 \\
%Q$\cdot$F$_p$      (arb. units)    &   8.7 (36)        &   2(37)       &       &   3.7(37)     &   1.8(36) &   4.7 (36)    &   2(36)   \\
\hline
\label{table:obs} 
\end{tabular}
\tablenotetext{a}{Shown is column 1 are: the \emph{Chandra} observation date, proposal number, and exposure time, T$_{exp}$ in kilo-seconds; comet-sun distance, r$_h$, comet-Earth distance, $\Delta$, neutral gas production rate, Q$_{gas}$, the solar wind proton velocity,  v$_p$ and proton flux, F$_p$ from the ACE-SWEPAM, SOHO-CELIAS on-line data archive.}
\tablenotetext{b}{Nomenclature and references: LS4=C/1999 S4 (\textsc{linear}), \citet{Lis01}; IZ=153P/2002  (Ikeya--Zhang), \citet{Bod07}; MH =C/1999 T1 (McNaught--Hartley), \citet{Kra02}; Encke=2P/2003 (Encke), \citet{Lis05}; 8P= C/2008 8P (Tuttle), \citet{Chr10}.} 
\end{table}
%\vspace{-0.15in}
%\end{minipage}
%\end{sidewaystable*}
\end{centering}

\subsection{Analysis}
The Chandra data reduction and analysis were done in the same manner as for our previous comet observations \citep{Lis01, Lis05, Bod07, W09, Chr10}
% (Lisse et al. 2001, 2005; Bodewits et al. 2007; Wolk et al. 2009; Christian et al. 2010) 
and the reader is referred to these papers for additional details. 
For each comet, a total image was created from combining each set of observations ({\it obsids}).  Individual images from these {\it obsids} were exposure corrected and remapped into a coordinate system moving with the comet using the {\it sso\_freeze} algorithm, part of the \emph{Chandra} Interactive Analysis of Observations (CIAO) software (Fruscione et al. 2006).  Spectra and associated calibration products (response matrices and effective areas) were extracted for each comet with the CIAO tools (CIAO v4.3) and analyzed with a combination of IDL and FTOOL packages.   
Spectral parameters were derived
using the least squares fitting technique with the \textsc{xspec} package \citep{Arn96}.

\subsubsection{Background Subtraction}
Due to the large extent of cometary X-ray emission, and
\emph{Chandra}'s relatively narrow field of view, it is not
trivial to obtain a background uncontaminated by the comet and
sufficiently close in time and viewing direction. Since our study was motivated by the strong excess emission observed for IZ in the 1300-2000 eV range, it is important to obtain good background subtraction, especially in the 1--2 keV region, where the background counts can be a significant fraction of the cometary source counts.

We tried several different methods for subtracting the background, including using an aperture for the outer region of the S3 chip, regions on the S1 chip, and ACIS observations of other sources at similar galactic latitudes. One problem with subtracting background from the S3 chip is the large physical extent of most comets and having comet photons in these outer regions. \citet{Bod07} used background extracted from the S1 chip, due to several very extended comets, such as Comet Q4 (NEAT). Using the S1 background has the advantage of being at the same time and same particle environment as the S3 observations, but may suffer slightly from poorer low energy response as compared to S3.  We also used a 450 ksec ACIS {\it blank sky}\footnote[2]{http://cxc.harvard.edu/contrib/maxim/acisbg} observation from the \emph{Chandra} archive.  We fit the IZ S3 spectrum with these various backgrounds, still noting important emission lines from the SWCX + higher energy lines ($>$ 1 keV). Shown in Figure 1 are several examples of the IZ spectrum with these different background spectra subtracted. We also summarize several of the important lines in Table 2. The ACIS blank sky observation rises more steeply than our S3 comet background and thus over-subtracts the spectrum above $\approx$1.7 keV. This background also has a strong feature at 1.74 keV, that is probably the result of the local, transient particle background \citep{Mark03} and we discount this line as real cometary emission. Lines near 1.35 keV are still present for the IZ spectrum when subtracting the ACIS black sky spectrum or even the S3 background raised to 3 times its nominal value. We find that the emission near 1.35 keV is signficant for the IZ spectra with all backgrounds, but the Si emission at 1.74 keV is suspect. We then  investigated  these 1.8-2.0 keV features using several calibration spectra (such as G~21.5$-$0.9;  \citet{Tsu11}, and the Crab; \citet{CS97}) observed with ACIS, and also concluded that these features were not related to calibration issues, such as the Si dead layer, and were true cometary or at least celestial emission.

%%\footnotetext{}{http://cxc.harvard.edu/contrib/maxim/acisbg}

No comets sampled for this study extended beyond the S3 chip and we therefore used background spectra from the S3 CCD extracted from rectangular areas near its outer edges at distances generally $>8\arcmin$. The subtracted background was then scaled by the ratio of its area relative to the source area.
 There are likely photons from the comet on the entire S3 chip, and some small iterations in the background scaling factor during fitting were performed as to not over-subtract the background. We feel using the S3 background spectrum at the same time and particle environment provides the most consistent spectral results. 
 In general our results for the 300-1000 eV range are in very good agreement with earlier works \citep{Lis01, Kra02} and the survey of  \citet{Bod07}
when our larger extraction area is taken into account. 

\subsubsection{Spectral Models}
We modeled our cometary spectra with a model based on the current SWCX model of \citet{Bod07} extended with individual emission lines to 2000 eV.
At first we had modeled the emission in the 280 to 1700 eV range, but then noticed significant emission in the 1700 to 2000 eV range.   We then expanded our analysis range up to 2000 eV.  In Figure \ref{fig:iz} we compare the IZ spectrum fit to 1300 eV to that fit to 2000 eV, showing the need for lines in the 1300 to 1700 eV range and that additional lines in the 1700 to 2000 eV range are also required. 

Our model includes the Bodewits et al. 2007 model, originally calculating line strengths from 300 to 1200 eV, now combined with lines above 1200 eV.  What follows is a brief description of the Bodewits et al. model which includes lines from C, N, O, and Ne. 
The \citet{Bod07} SWCX model follows the charge exchange
processes between solar wind ions and coma neutrals  both in the change of the ionization state of the solar wind ions and in the relaxation cascade of the excited ions.
Electron capture by highly charged ions populates highly excited
states, which subsequently decay to the ground state. These
cascading pathways follow the ionic branching ratios.
The absolute intensities of the emission lines are derived from a 3-D integration assuming cylindrical symmetry around the
comet--Sun axis.  Each group of ions in a species is fixed
according to their velocity dependent emission cross sections to the ion
with the highest cross section in that group. 
Thus, the free parameters of our model are the 
relative strengths of fully stripped- and H--like carbon (CV 299 and  C~VI 367 eV lines),
nitrogen (N VI 419 and NVII 500 eV) and oxygen (O~VII 561 and O~VII  653 eV) ions, 
%were free parameters for the spectral fitting. 
 lines for Ne IX at 922 eV and Ne X at 1023 eV and their weaker, but important transitions.  This model has %ur first model included the 
8 groups of fixed lines for C, N, O and Ne plus 6 lines added  for the 1200 to 2000 eV range at fixed energies. For our initial fitting, we first allowed the lines to vary within $\pm$30 eV, but found minimal improvement in $\chi^2$ and fixed the lines at their laboratory values.  All line widths were fixed at the \textsc{acis-S3} instrument resolution.
We note that the charge exchange cross sections for stripped- and H-like Mg and Si are currently lacking in the literature and for this reason we did not link any of the higher energy lines.  We used one line at 1340 eV to represent the helium like Mg XI forbidden (f), resonance (r), and intercombination (i) lines, and one line at 1850~eV characterizing helium like Si XIII forbidden (f), resonance (r), and intercombination (i) lines. We also included lines at 1470 (Mg XII  Ly $\alpha$), 1600 (Mg XI) , 1740 (Si K), and 1950 eV (Si XIV Ly $\alpha$).  

Thus, the spectra were fit in the 280 to 2000~eV range. This provided 118
spectral bins, and 104
degrees of freedom. 
\textsc{acis} spectra below $\approx$280~eV are discarded because of the rising background contributions, calibration problems and a decreased effective area
near the instrument's carbon edge (the ACIS energy map is only calculated down to 240 eV). 

%%%% new BG Table for IZ
 %_____________________________________________________________
\begin{deluxetable}{c c c c c c c}
\tabletypesize{\tiny}  % was small
\tablenum{2}
\tablecaption{IZ Fitted with the SWCX model and different backgrounds}
\tablehead{
\colhead{\textbf{$E_{line}$}}
& \colhead{} 
& \multicolumn{5}{c}{\textbf{Line Flux\tablenotemark{a}}} 
\\
\cline{3-7} 
%& \colhead{\textbf{Line} 
%& \colhead{\textbf{IZ}}
%& \colhead{\textbf{LS4}} 
%& \colhead{\textbf{MH}} 
%& \colhead{\textbf{Encke}} 
%& \colhead{\textbf{8P}} 
\colhead{\textbf{(eV)}}
& \colhead{\textbf{Line ID\tablenotemark{b}}}
& \colhead{\textbf{No BG}}
& \colhead{\textbf{S3 Bg}} 
& \colhead{\textbf{S1 BG}} 
& \colhead{\textbf{Blank Sky}} 
& \colhead{\textbf{S3 3xBG}} 
}
\startdata
299 & CV {\it f}+{\it r}+{\it i} & 
  4290$\pm$200   &
  3250$\pm$260 &
  3640$\pm$290    &
  2520$\pm$150 &
  1190$\pm$320
\\  
%3235$\pm$160 & 327$\pm$20 & 334$\pm$21 & 209$\pm$18 & 346$\pm$45 \\
%367.5 & CVI Ly$_{\alpha}$ & 420$\pm$40 & 71$\pm$7 & 54$\pm$6 & 24$\pm$4 & 23$\pm$6 \\
%419.8 & NVI $f+r+i$ & 187$\pm$28 & 41$\pm$3 & 22$\pm$3 & 11$\pm$2 & 13$\pm$3 \\
%500.3 & NVII Ly$_{\alpha}$ & 231$\pm$14 & 15$\pm$2 & 19$\pm$2 & 3$\pm$0.6 & 8$\pm$1 \\
561.1 & OVII {\it f}+{\it r}+{\it i} & 
 790$\pm$10&  % no BG 
 680$\pm$12 & % S3 Bg
 630$\pm$12 &    % S1 BG
 780$\pm$10  &   % Blank sjy
 450$\pm$30      % 3x bg 
\\     
%680$\pm$8 & 71$\pm$1 & 55$\pm$1 & 11$\pm$.4 & 19$\pm$0.5 \\
%653.5 & OVIII Ly$_{\alpha}$ & 727$\pm$6 & 39$\pm$1 & 41$\pm$1 & 5$\pm$0.2 & 7$\pm$3 \\
%907 & Ne IX $f+r+i$ & 52$\pm$1 & 5$\pm$0.3 & 4$\pm$0.1 & 1.0$\pm$0.1 & 1.3$\pm$0.1 \\
%1024 & Ne X & 17$\pm$1 & 7$\pm$2 & 2.0$\pm$0.2 & 1.0$\pm$0.1 & 2.0$\pm$0.1 \\
%1024 & Ne X & 17$\pm$1 & 78$\pm$5 & 2.0$\pm$0.2 & 1.0$\pm$0.1 & 2.0$\pm$0.1 \\
%1253 & Mg X K$_{\alpha}$ & 4.0$\pm$0.6 & 45$\pm$0.4 & 2.0$\pm$0.2 & 0.9$\pm$0.1 & 1.0$\pm$0.1 \\ 
1340 & Mg XI  {\it f}+{\it r}+{\it i} &
 30$\pm$4 &
 24$\pm$3 &
 23$\pm$5  &
 26$\pm$4 &
 13$\pm$3 
\\ 
% 22$\pm$7 & ... & $<$3 & ... & ... \\ % ??
%1351 & Mg XI {\it f} & ... & 5$\pm$4 & $<$3 & 0.9$\pm$0.1 & 0.7$\pm$0.1 \\
%1740 & CE Si K & 2.3$\pm$0.3 & 7.0$\pm$0.4 & 3.0$\pm$0.2  & 2.0$\pm$0.1  & 3.0$\pm$0.1  \\
1740 & Si K & 
3.1$\pm$0.3 &
2.0$\pm$0.5 &
0.8$\pm$0.6 &
 $<$1 &
% 0.7$\pm$1.0 
  $<$2.0 
 \\

1850 & Si XIII  {\it f}+{\it r}+{\it i} & 
7$\pm$0.5&
6$\pm$0.6 &
4$\pm$0.6 &
0.6$\pm$0.1 &
2.2$\pm$1 
\\
%5.0$\pm$0.4 & 6.0$\pm$0.3  & 2.0$\pm$0.2  & 1.0$\pm$0.1  & 2.0$\pm$0.1  \\
%1950 & Si XIV & 0.7$\pm$.2 & 8$\pm$2 & 2.0$\pm$0.8 & 2$\pm$0.4 & 0.9$\pm$0.5 \\
\hline
{\bf $\chi^2$/dof} &  &
2.2 &
1.2 &
1.8  &
5.7 &
0.9 
\\
%& $1.76$ & $1.7$ & $1.4$ & $2.4$ & $1.6$ \\
%\hline
\hline
%\bf{Flux\tablenotemark{c} $(0.30 - 1.0)$ keV} & & 34.4 & 3.23 & 2.66 & 0.75 & 1.07 \\
%\hline
{\bf Flux\tablenotemark{c} $(0.3 - 2.0)$ keV} & & 42.6 & 35.8 & 32.8  & 40.7 & 22.5   \\
\hline
\enddata
%\vspace{-0.2in}
%\footnotetext[1]{153P/2002;C/1999 S4;C/1999 T1;2P/2003;8P/Tuttle}
%\footnotetext[2]{Line fluxes 10$^{-5}$ ph cm$^{-2}$ s$^{-1}$}
\tablenotetext{a}{. Line fluxes 10$^{-5}$ ph cm$^{-2}$ s$^{-1}$}
\tablenotetext{b}{The forbidden (f), resonance (r), and intercombination (i) lines are included individually in our SWCX model \citep{Bod07}, except for Mg XI and Si VIII where they are combined as 1 line at 1340 eV, and 1854 respectively (see text).}
%\vspace{-0.2in}
\tablenotetext{c}{ in units of 10$^{-12}$ ergs cm$^{-2}$s$^{-1}$}

\end{deluxetable}

\section{Results}\label{sec:results}

\begin{figure}
   \centering
%          \includegraphics[width = 7.5cm]{rr_fig2log_ls4_spec.eps}
 %       \includegraphics[width = 7.5cm]{rr_fig2log_McH_spec.eps}
  %      \vspace{-0.4in}
     %       \includegraphics[width = 7.5cm]{rr_fig2log_enck_spec.eps}
       %     \includegraphics[width = 7.5cm]{rr_fig2log_8p_spec.eps}
                     \includegraphics[width = 7.5cm]{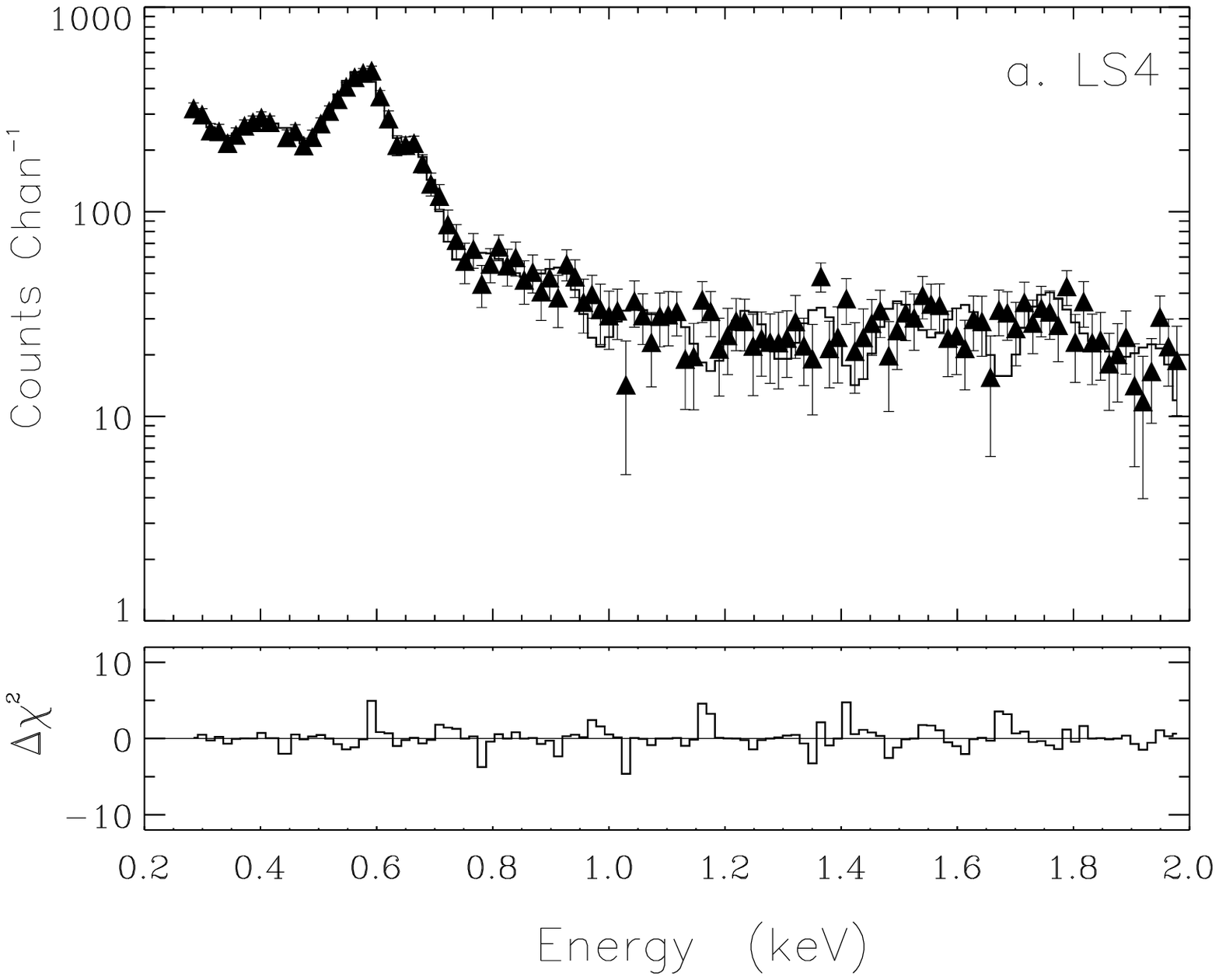}
        \includegraphics[width = 7.5cm]{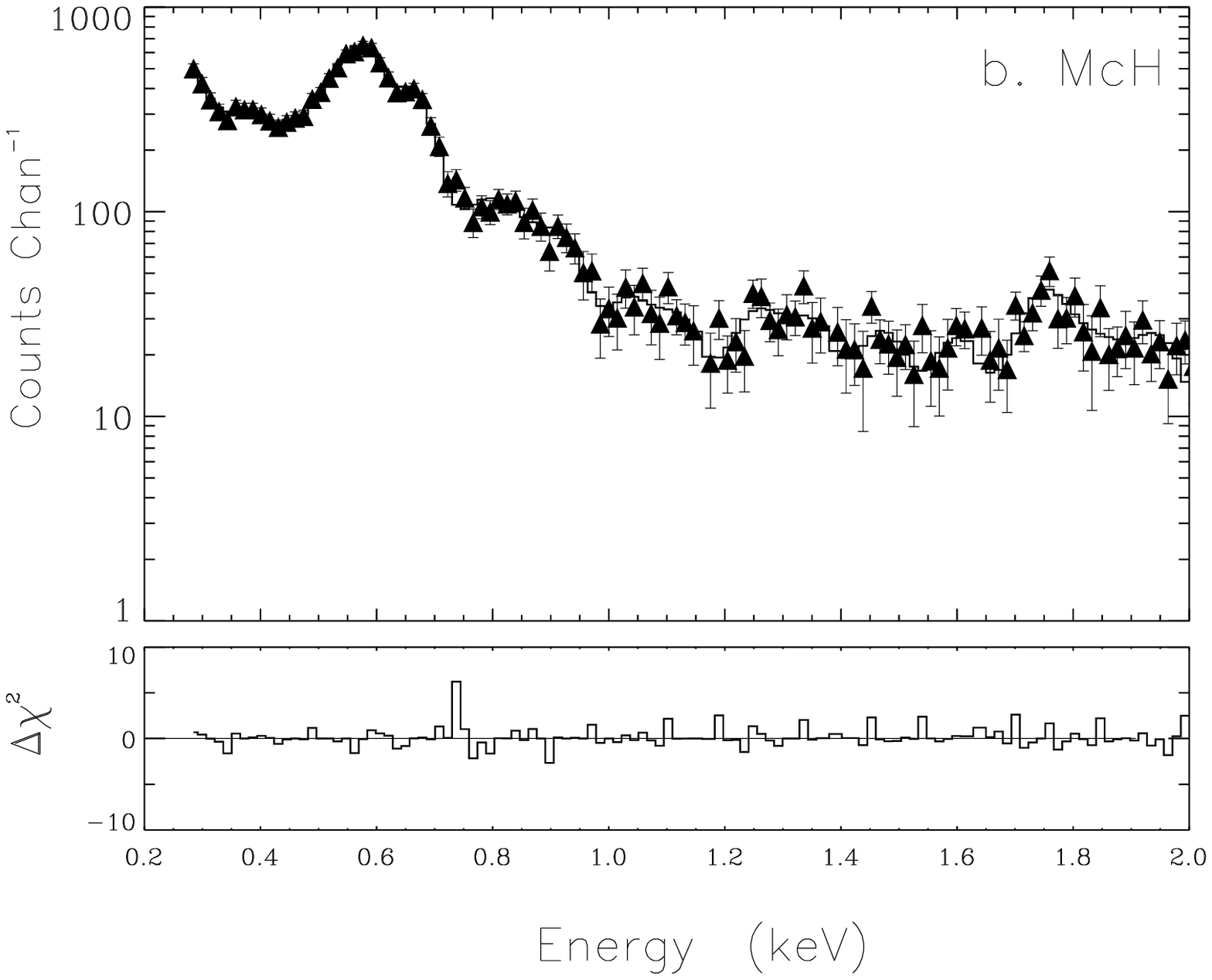}
        \vspace{-0.4in}
            \includegraphics[width = 7.5cm]{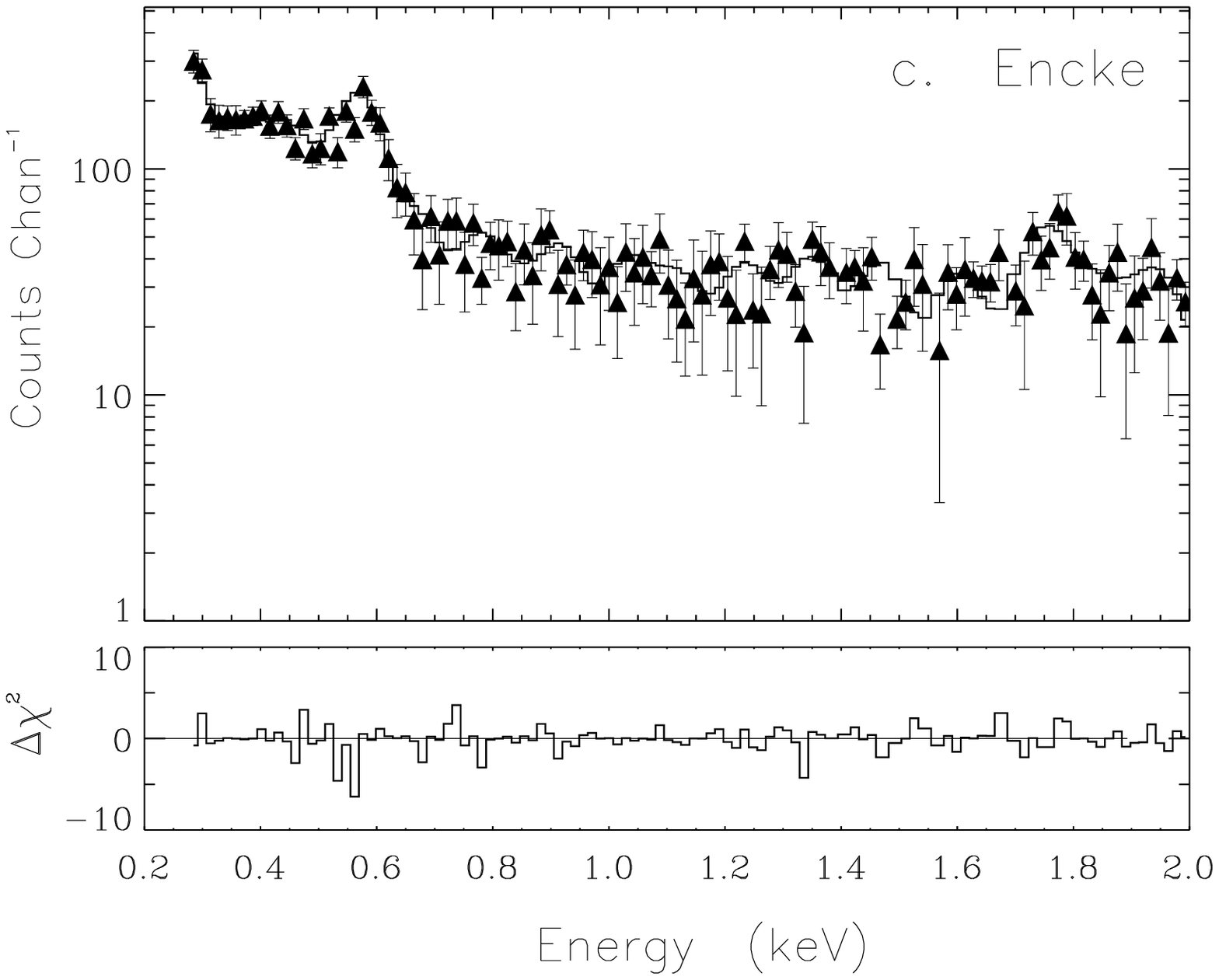}
            \includegraphics[width = 7.5cm]{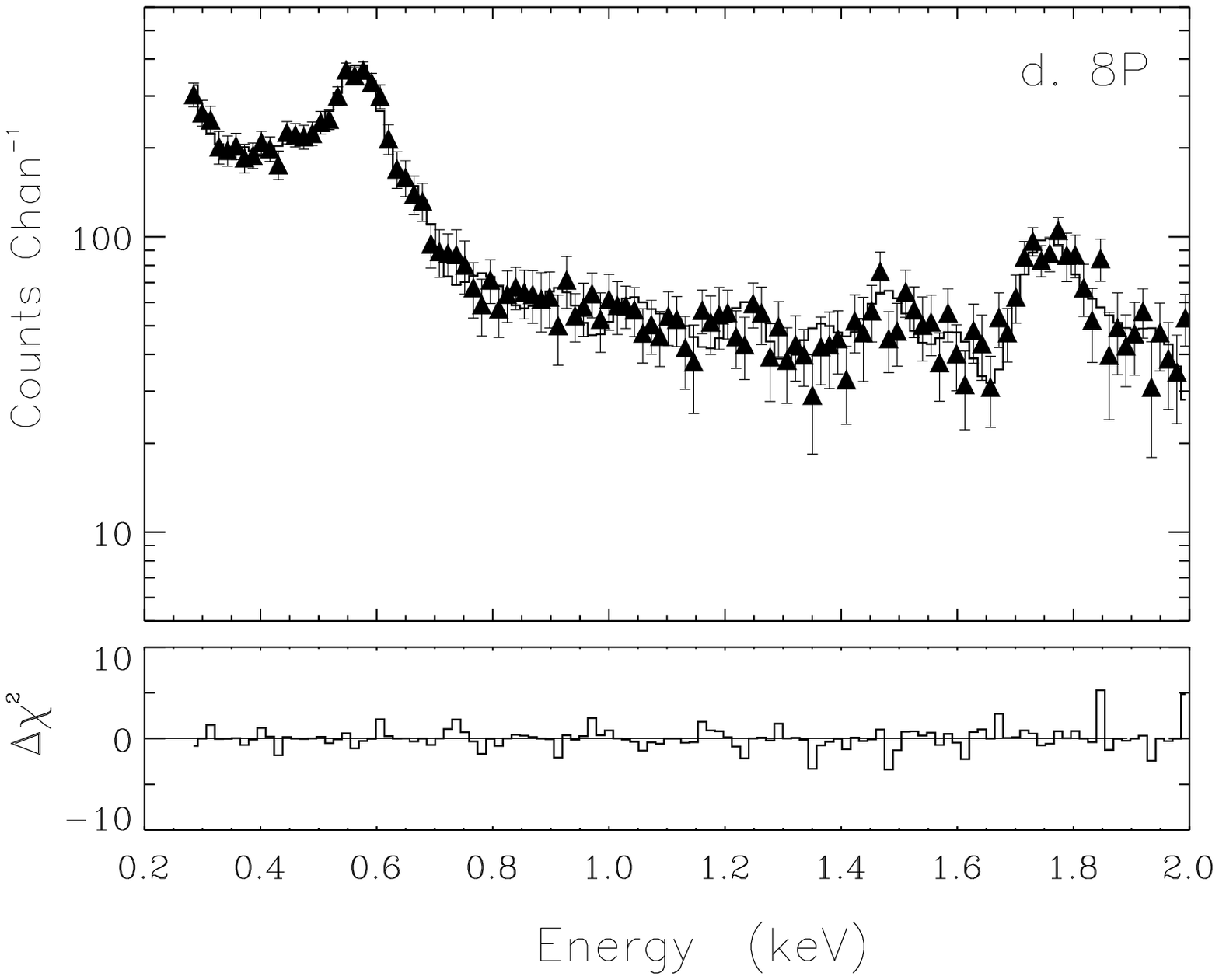}
   \caption{Observed spectra for our sample of comets fitted with the SWCX + higher energy line model.
   %multi-line model (Model 2). 
   The lower panel shown for each spectrum is the $\chi^2$ residuals of the data - model ($\Delta\chi^2$). The comets shown are: {\it a.} C/1999~S4~(\textsc{linear}, LS4), {\it b.} C/1999 T1~(McNaught--Hartley, MH), {\it c.} 2P/2003~(Encke),  and {\it d.} C/2008~8P~(Tuttle, 8P). (153P/2002~(Ikeya--Zhang), is shown in Figure 3.)
\label{fig:spec4}}
\end{figure}

Our study was motivated by the strong excess emission observed for Ikeya--Zhang  in the 1300-2000 eV range, and we confirm significant comet emission in this energy range for comet IZ and several additional comets. The Bodewits SWCX (Solar Wind Charge Exchange) model created for the 300 to 1200 eV range, extended to higher energies 
produces a $\chi^2$ for IZ when fitting the 280 to 2000 eV range of  $>$5000 (reduced $\chi^2$ $>$ 50). The addition of an emission line at 1340~eV reduces $\chi^2$ by 1362. Additional lines at 1470 and 1850 eV lower $\chi^2$ by about 200 each.  A line at 1740 eV reduced $\chi^2$ by 58 but we attribute this to a background feature (discussed below). The addition of a line at 1950 eV marginally reduced $\chi^2$ by only 7. 
Thus, the spectral fit with our SWCX model 
produced a $\chi^2$/dof  of 1.2 for IZ and this spectrum, residuals, and model are shown in Figure~3. 
The addition of  the 1340 eV feature to the spectra of the other comets (LS4, MH, 8P and Encke) reduced $\chi^2$ by 50 to 200 and the reduced $\chi^2$  for our SWCX model  range from 0.8 for 8P to 0.9 for MH and 1.23 for LS4.  Our SWCX model line fluxes and fit parameters are given in Table 3 and the reductions in $\chi^2$  for this model are given in Table 4. The spectral fits for our SWCX model are also shown in Figure \ref{fig:spec4} for LS4, MH, Encke, and 8P.  

 %_____________________________________________________________
   
%%%  log figure

%\section{Spectral Fitting}

 %_____________________________________________________________
\begin{deluxetable}{c c c c c c c}
\tabletypesize{\tiny}  % was small
\tablenum{3}
\tablecaption{SWCX Model Line Flux Results}
% new Aug 2012
\tablehead{
\colhead{\textbf{$E_{line}$}}
& \colhead{} 
& \multicolumn{5}{c}{\textbf{Line Flux\tablenotemark{a}}} 
\\
\cline{3-7} 
%& \colhead{\textbf{Line} 
%& \colhead{\textbf{IZ}}
%& \colhead{\textbf{LS4}} 
%& \colhead{\textbf{MH}} 
%& \colhead{\textbf{Encke}} 
%& \colhead{\textbf{8P}} 
\colhead{\textbf{(eV)}}
& \colhead{\textbf{Line ID\tablenotemark{b}}}
& \colhead{\textbf{IZ}}
& \colhead{\textbf{LS4}} 
& \colhead{\textbf{MH}} 
& \colhead{\textbf{Encke}} 
& \colhead{\textbf{8P}} 
}

\startdata
%265 & Mg IX, Si VIII, Si IX                         & 630$\pm$40 & 74$\pm$10  & 81$\pm$11 & 40$\pm$4  & 25$\pm$3    \\
299 & CV {\it f}+{\it r}+{\it i}                & 3250$\pm$260 & 330$\pm$50 & 340$\pm$50 & 220$\pm$60 & 430$\pm$90 \\
367.5 & CVI Ly$_{\alpha}$ & 414$\pm$70 & 66$\pm$20 & 54$\pm$14 & 17$\pm$11 & 25$\pm$12 \\
419.8 & NVI {\it f}+{\it r}+{\it i}          & 200$\pm$28 & 38$\pm$13 & 23$\pm$5 & 11$\pm$6 & 13$\pm$7 \\
500.3 & NVII Ly$_{\alpha}$ & 240$\pm$14 & 15$\pm$6 & 21$\pm$2 & 3$\pm$2 & 6$\pm$3 \\
561.1 & OVII {\it f}+{\it r}+{\it i}   & 680$\pm$12 & 73$\pm$4 & 56$\pm$1 & 12$\pm$2 & 20$\pm$2 \\
653.5 & OVIII Ly$_{\alpha}$ & 714$\pm$10 & 34$\pm$3 & 37$\pm$1 & 2$\pm$1 & 6$\pm$1 \\
907 & Ne IX {\it f}+{\it r}+{\it i}             & 53$\pm$2 & 5.5$\pm$0.7 & 4$\pm$0.5 & 1.0$\pm$0.2 & 1.3$\pm$0.2 \\
1024 & Ne X                       & 17$\pm$1 & 6$\pm$1 & 2.3$\pm$0.5 & 1.0$\pm$0.3 & 2.0$\pm$0.3 \\
%1024 & Ne X & 17$\pm$1 & 78$\pm$5 & 2.0$\pm$0.2 & 1.0$\pm$0.1 & 2.0$\pm$0.1 \\
%1253 & Mg X K$_{\alpha}$ & 4.0$\pm$0.6 & 45$\pm$0.4 & 2.0$\pm$0.2 & 0.9$\pm$0.1 & 1.0$\pm$0.1 \\ 
1340 & Mg XI  {\it f}+{\it r}+{\it i}    & 24$\pm$3 & 6.1$\pm$0.8 & 2.0$\pm$0.4 & 1.0$\pm$0.2  & 1.0$\pm$0.2 \\ % 
%1351 & Mg XI {\it f} & ... & 5$\pm$4 & $<$3 & 0.9$\pm$0.1 & 0.7$\pm$0.1 \\
1470 & Mg XII  Ly$_{\alpha}$               & 3.5$\pm$0.5 & 5.7$\pm$0.9 & 1.7$\pm$0.3 & 1.0$\pm$0.2  & 1.4 $\pm$0.2
\\
1600 & Mg XI & 2.1$\pm$0.6 & 7.2$\pm$0.4 & 1.5$\pm$0.5 & 1.0$\pm$0.2 & 1.0$\pm$0.3  
\\
1740 & Si K                  & 2.0$\pm$0.5 & 7.3$\pm$0.8 & 3.0$\pm$0.4  & 1.5$\pm$0.3  & 2.5$\pm$0.0.3  \\
1850 & Si XIII  {\it f}+{\it r}+{\it i}                   & 6.0$\pm$0.6 & 4.2$\pm$0.4  & 1.7$\pm$0.5 & 0.8$\pm$0.3  & 1.5$\pm$0.3  \\
1950 & Si XIV   Ly$_{\alpha}$            & 0.4$\pm$0.5 & 7.3$\pm$0.8 & 2.1$\pm$0.4 & 1.3$\pm$0.2 & 1.3$\pm$0.3 \\
\hline
{\bf $\chi^2$/dof} & & 1.18 & 1.23 & 0.93 & 1.04 & 0.80 \\
%\hline
\hline
{\bf Flux\tablenotemark{c} $(0.30 - 1.0)$ keV} & & 34.4 & 3.23 & 2.66 & 0.75 & 1.07 \\
%\hline
{\bf Flux\tablenotemark{c}  $(0.30 - 2.0)$ keV} & & 35.8 & 4.39 & 3.04 & 1.10 & 1.40 \\
\hline
\enddata
%\vspace{-0.2in}
%\footnotetext[1]{153P/2002;C/1999 S4;C/1999 T1;2P/2003;8P/Tuttle}
%\footnotetext[2]{Line fluxes 10$^{-5}$ ph cm$^{-2}$ s$^{-1}$}
\tablenotetext{a}{. Line fluxes in units of 10$^{-5}$ ph cm$^{-2}$ s$^{-1}$}
\tablenotetext{b}{The forbidden (f), resonance (r), and intercombination (i) lines are included individually in our SWCX model \citep{Bod07}, except for Mg XI and Si VIII where they are combined as 1 line at 1340 eV, and 1850 respectively (see text).}
%\vspace{-0.2in}
\tablenotetext{c}{in units of 10$^{-12}$ ergs cm$^{-2}$s$^{-1}$}
\end{deluxetable}

%%%%%%%%%% Table 

\begin{deluxetable}{lcccccr}
%\tabelwidth{0pt}
\tabletypesize{\tiny}
\tablenum{4}
\tablecaption{$\Delta\chi^2$ Results for SWCX + Higher Energy Line Model}
\tablehead{
\colhead{Region}
 &\colhead{E$_{line}$}
 &\colhead{Possible ID}
 &\colhead{Norm}
 &\colhead{$\Delta\chi^2$}
 &\colhead{$\frac{\chi^2}{dof}$}
\\
\colhead{}
 &\colhead{(eV)}
 &\colhead{}
 &\colhead{10$^{-5}$ph cm$^{-2}$ sec$^{-2}$}
 &\colhead{}
 &\colhead{}
}
\startdata
153P/2002 \\
%OLD & 1351 & Mg XI {\it f}\tablenotemark{a} & 13 & 489 & 5 \\
%  (IZ)           & 1835 & Si XIII & 4.6 & 199 & 2 \\  % Si K$_{\beta}$ ??
 %                  & 1739 & CE Si K\tablenotemark{b} & 2.2 & 166 & 2 \\
%		  & 1344 & Mg XI {\it r} & 6 & 18 & 0.2 \\
%		  & 1950 & Si XIV\tablenotemark{c} & 1 & 7 &  0.3 \\
(IZ) &1340 & Mg XI  {\it f}+{\it r}+{\it i}     & 24$\pm$3  & 1362 &  15.20  \\
&1470 & Mg XII Ly$_{\alpha}$   & 3.5$\pm$0.5 &  174 &  2.95 \\
&1600 & Mg XI & 2.1$\pm$0.6   &  67 & 1.9     \\
&1740 & Si K\tablenotemark{a} & 2.0$\pm$0.5      & 58 &  1.76 \\
&1850 & Si XIII {\it f}+{\it r}+{\it i} & 6.0$\pm$0.6   & 199 &   3.21 \\                 
&1950 & Si XIV Ly$_{\alpha}$    & 0.4$\pm$0.5    &  7  &   1.23 \\                
%1580,   67,  1.85 

\\
C/1999 S4 \\
%& 1739 & CE Si K & 7 & 327 & 3.6 \\
(LS4)	&	1340 & Mg XI  {\it f}+{\it r}+{\it i}      & 6.1$\pm$0.8  & 177 &  3.07  \\
%& 1950 & Si XIV & 7 & 175 & 1.9 \\
%		& 1835 & Si XIII & 4 & 159 & 1.6 \\
%		& 1351 & Mg XI {\it f} & 6 & 65 & 0.5 \\
		 
& 1470 & Mg XII Ly$_{\alpha}$   &  5.7$\pm$0.9 & 188 &  3.18 \\
& 1600 & Mg XI & 7.2$\pm$0.4  & 180 & 3.1 \\
& 1740 & Si K\tablenotemark{a} &  7.3$\pm$0.8 & 198 &  3.28 \\
& 1850 & Si XIII {\it f}+{\it r}+{\it i}   &  4.2$\pm$0.4 &  70 & 1.96 \\
&1950 & Si XIV Ly$_{\alpha}$     &   7.3$\pm$0.8 & 181 &  3.11 \\

\\
C/1999 T1 \\  
% & 1739 & CE Si K & 3 & 180 & 1.9 \\
% (MH)		& 1835 & Si XIII & 1 & 159 & 1 \\
%		& 1950 & Si XIV & 2 & 90 & 1 \\
%		& 1351 & Mg XI {\it f}  & 2 & 16 & 0.2 \\
 (MH) & 1340  & Mg XI  {\it f}+{\it r}+{\it i}   & 2.0$\pm$0.4 &    50 &   1.43 \\
& 1470 & Mg XII Ly$_{\alpha}$   &  1.7$\pm$0.3  &  59 &  1.52 \\
&1600 & Mg XI & 1.5$\pm$0.5 & 48 & 1.4 \\
& 1740 & Si K\tablenotemark{b} &  3.0$\pm$0.4  &143 & 2.36 \\
& 1850 & Si XIII {\it f}+{\it r}+{\it i}  &  1.7$\pm$0.5 & 42 & 1.34 \\
&1950 & Si XIV     &  2.1$\pm$0.4 &   94 &  1.87 \\
\\

2P/2003  \\ %& 1739 & CE Si K & 2 & 315 & 3.4 \\
	%& 1950 & Si XIV & 1 & 177 & 1.9 \\
		%& 1835 & Si XIII & 1 & 140 & 1.3 \\
		%& 1351 & Mg XI {\it f}  & 1 & 125 & 1.2 \\
(Encke)	&1340 & Mg XI  {\it f+i+r}  & 1.0$\pm$0.2  &  83 &  1.86 \\
& 1470 & Mg XII Ly$_{\alpha}$  & 1.0$\pm$0.2   &   78 &   1.81 \\
&1600 & Mg XI & 1.0$\pm$0.2 & 117 & 2.3 \\
& 1740 & Si K\tablenotemark{a} &  1.5$\pm$0.3 &  107 &  2.12 \\
& 1850 & Si XIII {\it f}+{\it r}+{\it i}  & 0.8$\pm$0.3  &   20 &   1.20 \\
 & 1950 & Si XIV Ly$_{\alpha}$    & 1.3$\pm$0.2  &   116 &   2.21 \\ 
\\
8P/Tuttle \\
%& 1739 & CE Si K & 2.4 & 390 & 4 \\
%(8P)	       & 1835 & Si XIII & 1.2 & 121 & 1.3 \\
%	       & 1950 & Si XIV & 2 & 100 & 1 \\
 %              & 1351 & Mg XI {\it f}  & 0.4 & 3 & $<$0.05 \\  
(8P) & 1340  & Mg XI  {\it f}+{\it r}+{\it i}  & 1.0$\pm$0.2 &  56 &  1.37  & \\
 &1470 & Mg XII Ly$_{\alpha}$ & 1.4$\pm$0.2  &  141 &  2.24   \\
&1600 & Mg XI & 1.0$\pm$0.3 & 78 & 1.6 \\
&1740 & Si K\tablenotemark{a} & 2.5$\pm$0.3 &  295 &  3.82  \\
& 1850 & Si XIII {\it f}+{\it r}+{\it i} & 1.5$\pm$0.3  & 69 & 1.51 \\
& 1950 & Si XIV Ly$_{\alpha}$  &  1.3$\pm$0.3  & 78 &  1.60 \\

\enddata
%\tablenotetext{a} { \citet{Tor07}}
\tablenotetext{a}{Not cometary emission}
%\tablenotetext{b}{ \citet{Ran08,Dju05}}
%\tablenotetext{b}{ \citet{Car10}}
\end{deluxetable}

\section{Discussion} \label{sec:discussion}
 %compare line and comet flues %%%%%%
  Our line fluxes are in good agreement with previous studies of this \emph{Chandra} sample of comets. Our line fluxes are consistent with \citet{Bod07} when our larger extracted area is taken into account. We extracted spectra from the entire chip, generally about a 1.5 to 2 times larger area than the 7.5$\arcmin$ circular aperture used in \citet{Bod07}. 
We computed X-ray luminosities in the 0.3 to 2.0 keV range from the fluxes for our SWCX model given in Table 2 and \emph{Chandra}-Earth distances given in Table 1. We find luminosities ranging from 2.0$\times$ 10$^{14}$ erg s$^{-1}$ for 8P and Encke to 1.7$\times$ 10$^{15}$ erg s$^{-1}$  for McNaught-Hartley and 2.3 and 2.5 $\times$ 10$^{16}$ erg s$^{-1}$  for LS4 and IZ, respectively.  These 
values are consistent with previous values in the literature given for the 0.3 to 1.0 keV range for LS4 \citep{Lis01}, Encke \citep{Lis05}, and 8P \citep{Chr10}. The luminosities are also consistent with those predicted by SWCX models  \citep{Den10}. % (Dennerl et al. 2010).  
The additional 1.0--2.0 keV flux contributes only 4\% of the total flux for IZ,  13\% of the total flux for McNaught-Hartley, but $\approx$20\% of the total flux for 8P and Encke, and nearly 30\% of the total flux for Linear S4. 
 
%% Line IDS %%Line IDs
We identify the newly detected feature at 1340 eV  as Mg XI \citep{Car10, Car11, Fuj07}.
%(Carter et al. 2009; 2011, Fujimoto et al. 2006).  
The ACIS S3 does not have the resolution to distinguish between the Mg~XI 1330 eV intercombination, the 1344 eV resonance, and 1351 eV forbidden lines, and we fit the spectra with a single line at 1340 eV. This line was significant for all comets in the sample with $\Delta\chi^2$ ranging from 50 for MH to over 1300 for IZ. We also detect the Mg XII 1470 eV line at a significant level and the 1600 eV line of Mg XII is marginally detected.
%the 1351 eV line represents a $^1$S$_0$-$^3$S$_1$ forbiden line transmission \citep{Tor07}; we therefore expect it to contribute to the charge exchange model. Only IZ shows a significant additional line at  transition.
%{\it will have to double check with tables}...
\citet{Car10} report Mg XI lines at 1.34, 1.6 keV for the Earth's exosphere and Mg XII at 1.47 keV.
Additionally, for comets, 
Mg was reported in the low energy spectrum of Comet McNaught--Hartley in \citet{Kra02} near 250 eV, near the low energy cut-off of the ACIS-S3 CCD, and Mg ions have solar winds abundances similar to carbon and oxygen. However, searching for low energy emission was beyond the scope of the current work. 
% Djur\'ic et al. 2005 
We have detected a strong feature at 1740~keV for all comets.
 \citet{Dju05} identify the 1740 eV line with charge exchange from neutral silicon K to L and K to M transitions (Si $K-L_{2,3}$,$K-M_{2,3}$), but \citet{Car10} identify a feature at 1730 eV as Al~XIII. \citet{Ran08} also find a similar feature (near 1740 eV,  7.13 \AA) in the XMM spectrum of starburst galaxy M82 and also conclude it is charge exchange from neutral silicon.  However, neutral silicon would have to penetrate to the cometary surface in order for this line to possibly undergo charge exchange, and this line is one of the stronger features in the ACIS blank-sky observation (Figure 1). We therefore consider it to be a background feature and not cometary charge exchange emission. 
 %Based on the much lower Al solar wind abundance, we feel the Si identification is more likely, and we note this as {\it CE Si K} in Table 3.
Additional features found at 1850 and 1950 eV are identified as 
Si XIII and  Si XIV, respectively \citep{Car10}.  %(Carter 2010). 
%1.73 Al XIII (Carter 2009).
The 1s2s$^1$S$_0$--1s2p$^3$P$_1$ intercombination interval was measured for helium like silicon by \citet{RM02} producing 1850 eV X-rays, suggesting this may contribute to the charge exchange model as well. Once again due to the limitations of ACIS resolution, we must consider the Si XIII line at 1850 eV to contain forbidden line at 1838~eV the 1854~eV intercombination line, and the resonance line at 1863~eV. The    Si~XIV line at 1950~eV is only marginally detected for IZ, but is significant for the other comets. 

%  line ratios. 
Calculating flux ratios of observed emission lines or grouping of these lines is useful for determining relative ionic abundances and hence the solar wind state and conditions in the comet's nucleus for the optically thin case. These ratios may then be used to test predictions of SWCX models and in separating cometary X-ray spectra into classes \citep{Bod07}.  In general carbon and oxygen
are the most abundant heavy ions in the solar wind and have produced some of the strongest features reported from cometary X-ray spectra \citep{Cra97, SC00, KD00, Kra06}. Thus, 
the ratio of the sum of the lower energy lines ($\approx$ 300--500 eV) to the strongest feature, OVII is a very sensitive diagnostic to the solar wind conditions (See Figure 6 in \citet{Bod07}).  In Figure 5ab we show several line flux ratios as a function of both the OVIII/OVII flux ratio and solar wind velocity. 
Similar to \citet{Bod07} we find the Carbon + Nitrogen to OVII line ratio (C+N/OVII) to be anti-correlated with the OVIII/OVII ratio. For cold and fast solar wind most of the oxygen is in O$^{6+}$ and thus does not contribute to SWCX emission in the X-ray portion of the spectrum and the C+N/OVII ratio is larger. Mg and Si can have significant abundances in the typical solar wind \citep{vss00, zf02}. % removed 15\%
%are found to be about one-third to one-half of Oxygen for the xx solar wind (. 
Our higher energy lines of Mg XI and Si  line ratios also show a similar trend to the C+N/OVII ratios, decreasing with an increase in the OVIII/OVII ratio.  
% - New Sept 18 2012
We find MgXI/OVII ratios ranging from 0.04 for IZ to 0.17 for LS4. These are similar to the MgXI/OVII ratio of 0.04 found for the SWCX contribution of the X-ray background \citep{Sno04} and 0.14 found by \citet{Fuj07} for the Earth's magnetosheath, but much lower than the ratio of 0.28 that  \citet{Car10} found for the Earth's exosphere after a coronal mass ejection (CME) event. Our measured Si XIII/OVII ratios range from 0.01 for IZ to 0.08 for 8P and similar to the Mg ratios, but they are smaller than the ratio of 0.30 found by \citet{Car10}.

%Coronal mass ejections 
CMEs and interplanetary CME's (ICME's) can cause unusually high charge states of C, O, Ne, Si, Mg, and Fe in the solar wind \citep{Sno04, Gru11}. 
\citet{Gru11} show that silicon primarily has charge states of Si$^{8+}$ to Si$^{9+}$ for a typical solar wind and this is shifted to  Si$^{10+}$ and Si$^{11+}$
for ICME plasma solar wind and their time-of-flight instruments are limited to charge states between 8+ and 12+. Our Chandra results confirm moderate to strong Si emission for all comets in our sample from charge states of Si$^{13+}$ to Si$^{14+}$, but with the limited resolution and exposure times of most Chandra observations, a deep observation of a very active and X-ray bright comet is needed to search for additional charge states of Si. 
%\citet{Che03} show that Mg$^{10+}$ can contribute 30 to 51\% of ion fraction for fast SW...

No strong trends are seen for these line ratios as a function of solar wind velocity, but as discussed in \citet{Bod07} the velocity alone does not give a good indication of the solar wind state and underlying ioinic composition.  Additionally,  above $\approx$300 km s$^{-1}$, the charge exchange cross sections change only slightly with SW velocity \citep{SC00, Lis05}. However, the similar trends for Mg and Si ratios observed for the C+N ratios promise these features can be used for future diagnostics.

\begin{figure}
   \centering
       \includegraphics[width = 7.5cm]{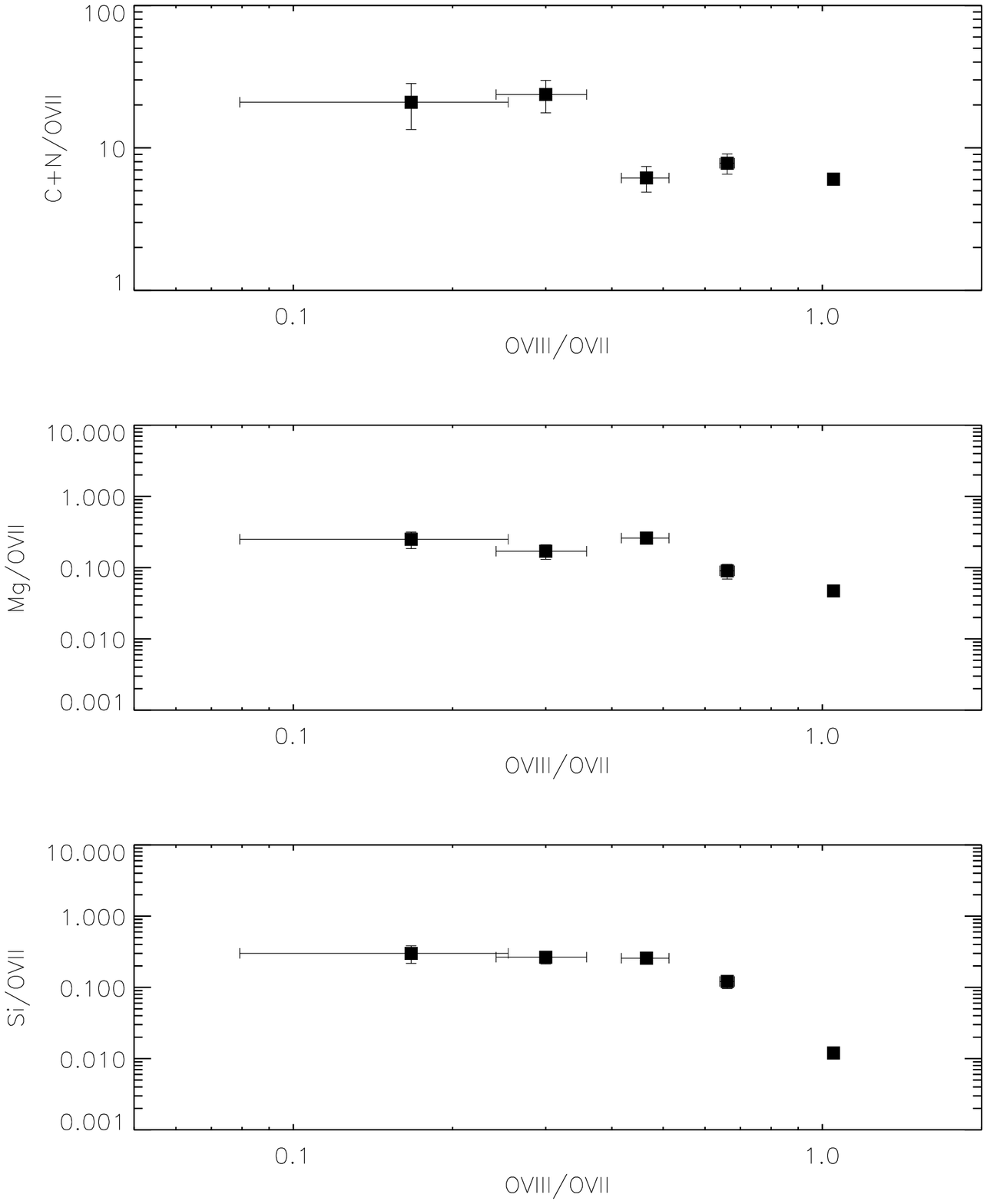}
       \includegraphics[width=7.5cm]{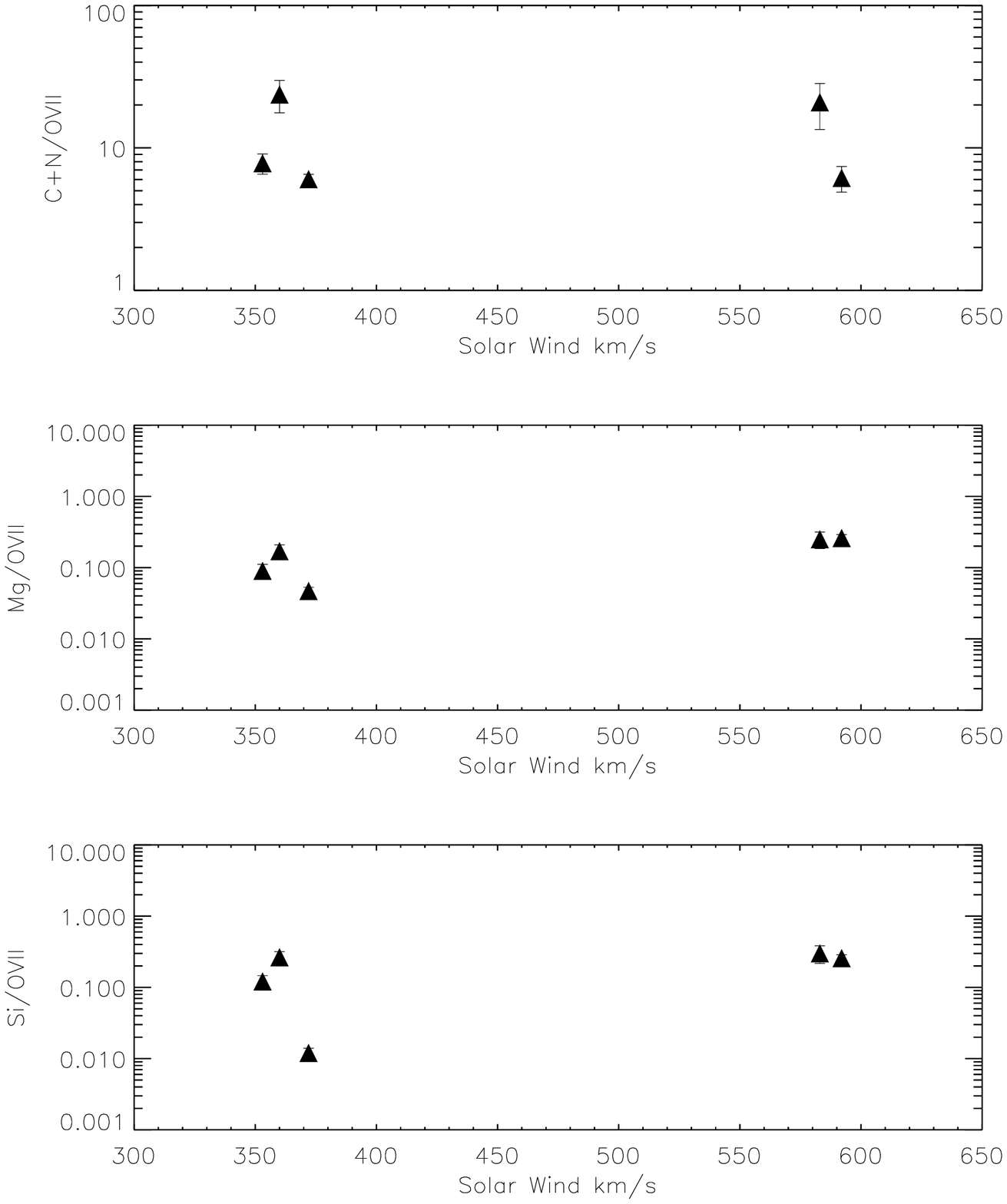}
   \caption{Flux ratios for our sample comets.  Shown in the left panel are flux ratios as a function of the OVIII/OVII ratio for the sum of the strongest carbon + nitrogen lines (C+N, top), Mg lines (middle) and Si (bottom panel). The right panel shows the same ratios as  a function of solar wind velocity (see text).
   %The low energy C+N feature is anti-correlated to the oxygen ionic ratio.
   % with \emph{Chandra}, grouped by their spectral shape (see text). The histogram lines indicate the SWCX model fit.
\label{fig:panel}}
\end{figure}
%______________________________________________________________________
\section{Conclusions}\label{sec:conclusions}
We have detected new line emission in the 1300-2000 eV range for several comets observed with \emph{Chandra} ACIS spectrometer.  The typical, previously reported emission lines in the 300 to 1000 eV range are detected for all comets from CV, CVI, NVI, NVII, OVII, OVIII, Ne IX, Ne X, and line strengths correlate with the charge state of the solar wind continuing credence for the idea that these lines are formed by solar wind charge exchange. 
%Above 1 keV, we find Ikeya--Zhang to have  strong emission lines at 1340, 1470, 1600 and 1850 eV that we identify as being created by solar wind charge exchange with Mg XI, Mg XII, Mg XI, and Si XIII, respectively. 
Above 1 keV, we find Ikeya--Zhang to have strong emission lines at 1340 and 1850 eV that we identify as being created by solar wind charge exchange lines of Mg XI and Si XIII, respectively, and weaker emission lines at 1470, 1600, and 1950 eV formed by SWCX of Mg XII, Mg XI, and Si~XIV, respectively. 
 The Mg XI 1340 line is significant for the other comets in our sample (LS4, MH, Encke, 8P).
%,  but only marginally detected for comet 8P. T
Additional Mg lines in the 1400 to 1600 eV range are detected for all comets in our sample at a significant level and promise additional diagnostic lines to be added to SWCX models.  
The Si XIII line at 1850 eV is detected for all comets in our sample. The Si XIV is only marginally detected for IZ, but is signifcant for the other comets in our sample.  
Although the silicon lines in the 1700 to 2000 eV range are detected for all comets, with the rising background and decreasing cometary emission, we caution these detections need further confirmation with higher resolution instruments. 
 % in our sample at a significant level and promise additional diagnostic lines to be added to SWCX models.  
%
%Mg Further investigation into emission $\sim$1600 eV is req	uired. \cite{Kat11} measure(claim) Mg He$\beta$ line emissions around 1.58 keV with \emph{XMM Newton} .... Residuals likely caused by \emph {Chandra's} aluminum ACIS filters? 

%__________________________________________________________________

\begin{acknowledgements}
This research was supported by \emph{Chandra} archival grants CXO-09100455 and CXO-13100089.
DC also thanks the CSUN Department of Physics and Astronomy for start-up funds in support of this research. We thanks S. Lepri for useful discussion on solar wind abundances, and we thank an anonymous referee for suggested improvements.
%We are grateful for the cometary ephemerides of
%D.~K.~Yeomans published at the \textsc{jpl/horizons} website.
%Proton velocities used here are courtesy of the
%\textsc{soho/celias/pm} team. \textsc{soho} is a mission of
%international cooperation between \textsc{esa} and \textsc{nasa}. 
\end{acknowledgements}

\end{document}